# Mechanism of the double heterostructure TiO$_2$/ZnO/TiO$_2$ for photocatalytic and photovoltaic applications: A theoretical study

**Slimane Haffad[1,2*]**

1) Department of Technology, Faculty of Technology, University A. Mira of Bejaia, Route Targa-Ouzemour 06000, Bejaia, Algeria.

2) Laboratory of Organic Materials, , Faculty of Technology, University A. Mira of Bejaia, Route Targa-Ouzemour 06000, Bejaia, Algeria.

**Abstract.** Understanding the mechanism of the heterojunction is an important step towards controllable and tunable interfaces for photocatalytic and photovoltaic based devices. To this aim, we propose a thorough study of a double heterostructure system consisting of two semiconductors with large band gap, namely, wurtzite ZnO and anatase TiO$_2$. We demonstrate via first-principle calculations two stable configurations of ZnO/TiO$_2$ interfaces. Our structural analysis provides a key information on the nature of the complex interface and lattice distortions occurring when combining these materials. The study of the electronic properties of the sandwich nanostructure TiO$_2$/ZnO/TiO$_2$ reveals that conduction band arises mainly from Ti$_{3d}$ orbitals, while valence band is maintained by O$_{2p}$ of ZnO, and that the trapped states within the gap region frequent in single heterostructure are substantially reduced in the double interface system. Moreover, our work explains the origin of certain optical transitions observed in the experimental studies. Unexpectedly, as a consequence of different bond distortions, the results on the band alignments show electron accumulation in the left shell of TiO$_2$ rather than the right one. Such behavior provides more choice for the sensitization and functionalization of TiO$_2$ surfaces.

## 1. Introduction

Solar energy conversion consists on the production of electrical energy in the form of current and voltage from electromagnetic energy: i.e., incident light including infrared, visible, and ultraviolet (UV) [1, 3, 2, 4, 5, 6]. The first generation solar cells were made of semi-conducting $p$–$n$ junctions (based on Si wafers) and the second generation is based on the improvement of the first generation by employing thin film technologies [3, 2, 1].

In the last years, a third generation has emerged, which includes non-semiconductor technologies (polymer cells and biomimetics) [2], nanowires (NWs) and quantum dots (QDs) [3]. Within the third generation, dye sensitized solar cells (DSSCs) as low-cost solar cell, clean, and renewable energy sources became a practical root for photovoltaic cells when Regan and Graetzel [4], in the late 1980s, have fabricated DSSCs composed of a porous layer of titanium dioxide ($TiO_2$) nanoparticles immersed under an electrolyte solution and, covered with a dye molecule that absorbs sunlight. Such a technique expanded the use of semiconductors with wide band gap such as GaN, $SnO_2$, SiC, and ZnO which makes possible the conversion of higher energy photons. Ten years before this invention, Fujishima and K. Honda [5] discovered the effect of photosensitization of the $TiO_2$ electrode under UV irradiation for the photocatalytic water splitting.

It is well known that interfacial charge recombination is a serious problem for photocatalytic and photovoltaic based devices [1, 3, 2, 6]. Such a phenomena causes a loss of photo-generated electrons. It affects the open circuit voltage by decreasing the concentration of electrons in the conduction band of the semiconductor and, also the photo-current by decreasing the forward injection current. From this point of view, nanostructures, in regard of their large surface to volume ratio, present an inconvenience, i.e., by increasing the probability of charge recombination. An attempt to reduce the recombination rate consists of using a bilayer of a metal-oxide semiconductors electrode for high-performance nanomaterial-based DSSCs. One of the proposed systems is core-shell structures, which are derived from the nanoparticles and can reduce the charge recombination by forming a coating layer. An established electric field that may assist the separation of the electrons in the solid-solid interface can form energy barriers at the electrode-electrolyte interface. For this reason, several materials have been tested such as $SnO_2/TiO_2$ [7, 8], $SnO_2/ZnO$ [9], $TiO_2/Nb_2O_5$ [10], $ZnO/Al_2O_3$ [11], and $ZnO/TiO_2$ [11, 12]. In the case of ZnO and $TiO_2$, their similar photovoltaic performances did not come from similarity in properties but from compensating ones [13]. Matt Law *et al.* [11] demonstrated the superior performance of the $ZnO/TiO_2$ core-shell nanowire (CS-NWs) cells if compared to $ZnO/Al_2O_3$ CS-NWs cells. Core- shell nanorod arrays based $ZnO/TiO_2$ encased in the hole-conducting polymer P3HT were performed and, a significant increases in the voltage and fill factor relative to devices without shells was observed [14]. In particular, Greene *et al.*[14] found that the shell-thickness affects the cell performance and, they showed that, adding a ∼ 5 nm polycrystalline $TiO_2$ shell improved the efficiency of the devices, while, Cr-doped $TiO_2$ nanoshell coating single-crystalline ZnO nanowires allows formation of *p-n* junctions via an efficient charge separation [15]. Park *et al.*[16] have realized photoelectrodes made of submicrometer-sized aggregates of ZnO nanocrystallites coated with $TiO_2$ layer by atomic layer deposition. They demonstrated that surface diffusion of the ZnO atoms at elevated annealing temperature can be suppressed and, the efficiency of DSSCs was enhanced with more than 30%. The effect of the interface $ZnO/TiO_2$ was also tested on the performance of polymer solar cells. It has been found that ZnO nanorod coated with $TiO_2$ layer demonstrates a significant reduction of the recombination rate and that the $TiO_2$ interface layer functions as an efficient photo-generated exciton quencher and assisted charge collection [17]. $ZnO/TiO_2$ hybrid nanostructures demonstrate also a higher catalytic activity [18], where an enhanced charge transfer/separation process with fine interfaces was observed.

Understanding the mechanism of ZnO/TiO$_2$ heterojunction and how the physical characteristics are affected by the solid-solid reaction are crucial for a better use in photovoltaic or photocatalysis systems. Unfortunately, considering the lack of experimental data on the mechanism of the interface between ZnO and TiO$_2$, we have based upon our own theoretical approach. However, some experimental studies were critical for the validation of our method. For exemple, Panigrahi and Basak [19] have found, using high resolution TEM (HRTEM) image, that the lattice fringes of ZnO/TiO$_2$ interface correspond to the (0002) plane of the wurtzite ZnO and to (112) plane of anatase (101) surfaces. One can draw similar observations from the works done by Wang *et al.* [20] and Greene *et al.* [14]. On the simulation level, few works have been done in this perspective. Conesa [21] reported that conduction band of wurtzite ZnO is more negative than anatase TiO$_2$ for TiO2/ZnO interface. While Meysam *et al.* [22] found rutile TiO2 shell changes the surface dipole distribution of ZnO nanowire causing a shift in the conduction band (CB) and valence band (VB) of ZnO to higher energies.

In this work, we report electronic structure calculations of pseudo-realistic ZnO/TiO$_2$ interfaces using density functional theory (DFT) and the double-macroscopic average technique. The paper is organized as follows. In Section 2, a systematic study on the stability of the heterostructure is presented within the appropriate computational approach. Section 3 is devoted to the results and discussions. We first analyze and discuss the effect of the distortions induced at the interfaces owing to the misfit dislocations, in term of atomic bonding and relaxation. Next, we calculate the electronic properties of the sandwich system TiO$_2$/ZnO/TiO$_2$, in which the resultls are argued and compared to other available works. Afterward, we examine the alignment of the energy levels around valence and conduction bands by mean of an accurate method based on the average of the electrostatic potential [25]. At the end, a summary of the main results is presented in Sec. 4. Our study provides key information on the nature of the interfaces when matching together wurtzite ZnO with anatase TiO$_2$ and its impact on the energy band offsets.

## 2. Model and method

The appropriate approach used to determine the stable interfaces is based on *ab initio* calculations where different orientations of ZnO/TiO$_2$ heterostructure are combined. Previous theoretical and experimental studies evidence (10$\bar{1}$0) and (101) facets like the most stable surfaces for ZnO [26] and TiO$_2$ [27, 28], respectively. Consequently, when bringing these two surfaces together, the interface will be: ZnO (10$\bar{1}$0) ∥ TiO$_2$ (101) or TiO$_2$ (10$\bar{1}$), that is, the mostly observed in experiment [11, 14, 19, 20]. However, in literature published so far, few information on the nature of the bonding and the presence or not of core dislocations were found. Nevertheless, It has been often noticed that ZnO core consists of a crystalline arrangement of atoms (hexagonal wurtzite structure), while TiO$_2$ layers seem to be formed by polycrystalline and porous regions with anatase phase.

The ZnO (10$\bar{1}$0) non-polar surface is defined by the following lattice parameters: $a_{ZnO}$ (∥Y) ≈ 3.23 Å, and $c_{ZnO}$ (∥Z) ≈ 5.27 Å [see Fig. 1(a)], whereas the TiO$_2$ (10$\bar{1}$) surface, also non-polar, is built by $a_{TiO2}$ (∥Y) ≈ 3.78 Å, and $c_{TiO2}$—[101] (∥Z) ≈ 10.32 Å ‡. Based on the experimental observations [14, 19] and structural information, we built our interface model by considering six (five) and two (one) unit cells in Y and Z directions, respectively, for ZnO (TiO$_2$) slabs. These parameters will be referred to as: $A_{ZnO} = 6a_{ZnO} \| A_{TiO2} = 5a_{TiO2}$, $C_{ZnO} = 2c_{ZnO} \| C_{TiO2} = c_{TiO2}[101]$. A combination of these two surfaces gives a reasonable lattice misfit less than 3%, with: $(\Delta C/C_{TiO2}) \approx (\Delta C/C_{ZnO}) \sim \pm 2.0\text{-}2.2\%$ and $(\Delta A/A_{TiO2}) \approx (\Delta A/A_{ZnO}) \sim \pm 2.7\text{-}2.8$. To reproduce the experimental situation, we considered a periodically repeated multilayer: the interface was built by matching (6 × 2) layers of ZnO (10$\bar{1}$0) with (5 × 1) layers of TiO$_2$ (101). On the TiO$_{2-[101]}$ ∥ ZnO$_{[0001]}$ interface direction (in Z direction as labeled in Fig. 1), the TiO2 super cell consists of eight parallel atomic layers of O and Ti § that matches with ZnO[0001] of eight parallel atomic layers (four of both O and Zn). Such a combination leads to an accordance with experimental observations deduced from the HRTEM image of Wang *et al.* [20] in which no core dislocation has been counted in this interface direction.

On the TiO$_{2-[010]}$ ∥ ZnO$_{[11\bar{2}0]}$ interface direction, the combination results in 10 and 12 slabs represented by the lattice parameters $A_{TiO2}$ and $A_{ZnO}$, respectively * (in Y direction as labeled in Fig. 1). This give rise to an edge dislocation in the supercell [Two equivalent dislocations lie at two different planes due to the very corrugated and sawtooth profile of the (101) surface of anatase TiO$_2$, see Ref. [27]]. Several tests were carried out in order to find and localize the stable configurations. Calculation tests were made by fixing the positions of ZnO (TiO$_2$) atoms and displacing the coordinates of TiO$_2$ (ZnO) atoms in both Z and Y directions with 0.3 Å until they coincide with the equivalent positions. We first moved the TiO$_2$ (ZnO) atoms in Y direction until the positions for which the minimum energy is reached, and then, by fixing the coordinates along Y at the minimal energy positions, the same procedure is followed for the Z direction.

The results on the interface orientations moved in Z directions are reported in Table 1. From values of the energies, we identified two most stable orientations for which the minimum energy is obtained. To get the unrelaxed double interface system

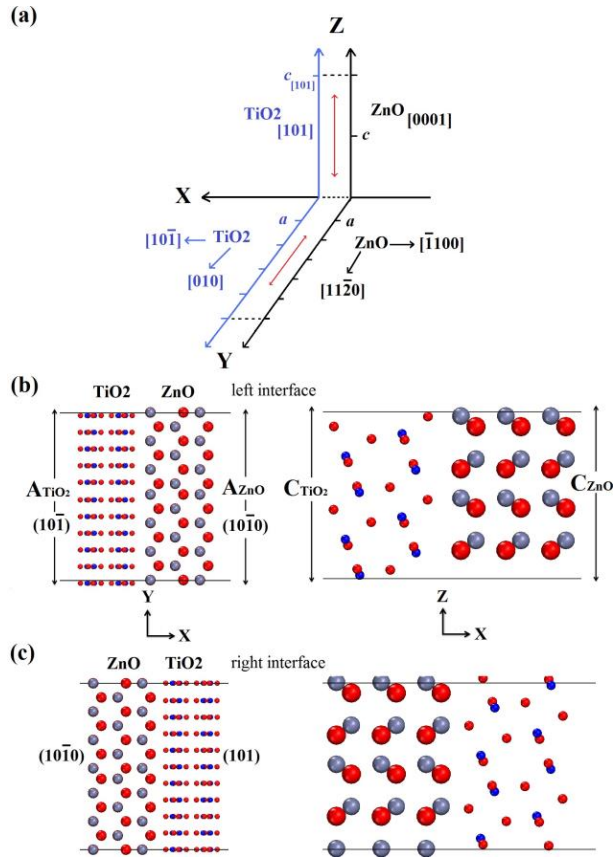

**Figure 1.** Schematic representation in 3-D system of ZnO/TiO$_2$ interface (a), unrelaxed heterostructure of the left interface TiO$_2$/ZnO (b), unrelaxed heterostructure of the right interface ZnO/TiO$_2$ (c). Oxygen, zinc, and titanium are represented by red, gray, and blue balls, respectively.

TiO$_2$/ZnO/TiO$_2$, we combined these two stables configurations in some sort of sandwich system in such a way that ZnO coated uniformly by TiO$_2$ slabs: the two outer ZnO sides enclosed by (101) or (10$\bar{1}$) surfaces of TiO$_2$ (see Figs. 1 and 2). Our double heterostructure has a slab thickness of ~ 28 Å (~ 15.7 Å of ZnO thickness and ~ 2 × 6 Å for TiO$_2$). By keeping the stoichiometry, the basic unit cell contains 528 atoms, {(ZnO)$_{144}$ and (TiO$_2$)$_{80}$}, periodically repeated in space within cubic boundary conditions separated by a vacuum region > 15 Å wide in $X$ direction.

We employed the density functional theory (DFT) [23, 24] method to investigate the properties of the sandwich system TiO$_2$/ZnO/TiO$_2$. Geometry relaxations are calculated within the generalized gradient approximation (GGA) as parameterized by Perdew, Burke, and Ernzerhof (PBE) [29]. The Kohn-Sham orbitals were expanded in numerical pseudo atomic localized basis sets (SIESTA package [30]) with double zeta polarization (DZP) and electron-ion interaction was included by employing norm-conserving pseudopotentials [31]. We used the zinc and oxygen pseudopotentials described elsewhere [32], whereas a relativistic pseudopotential for the ionic titanium including non-linear core corrections was generated using the code ATOM [33] in the following reference configuration: $3s^2\,3p^6\,3d^2\,4f^0$, with a cutoff radii of 1.5 a.u for $3s$ and $3d$, 1.4 a.u for $3p$, and 2.0 a.u for $4f$. The states $3s$ and $3p$ of titanium were treated as semicore levels, whereas $4s$ with $3d$ were taken as higher valence states (12 valence electrons including the ionic charge). The Brillouin zone (BZ) sampling was performed using (16×16×16) and (1×2×4) Monkhorst-Pack grid [34] for bulk and slab models respectively, and a mesh cutoff of 500 Ry is considered in a real space grid. Structural optimizations were performed using the conjugate gradient method and convergence was assumed when the atomic forces were less than 0.03 eV.

**Table 1.** The relative energies, $E - E_{min}$, of the different configurations computed for various displacements, $d$, with respect to Z axis. Interface TiO$_2$/ZnO; TiO$_2$ slabs lie at the left side of ZnO, and interface ZnO/TiO$_2$; TiO$_2$ slabs lie at the right side of ZnO. The star indicates the most stable configurations

| TiO$_2$/ZnO | | ZnO/TiO$_2$ | |
| --- | --- | --- | --- |
| $d\,(\updownarrow z)$ (Å) | $E - E_{min}$ (eV) | $d\,(\updownarrow z)$ (Å) | $E - E_{min}$ (eV) |
| -0.9 | 1.36 | -0.9 | 2.42 |
| -0.6 | 0.67 | -0.6 | 0.09 |
| -0.3 | 0.02 | -0.3 (⋆) | 0.00 |
| 0.0 (⋆) | 0.00 | 0.0 | 1.95 |
| 0.3 | 0.37 | 0.3 | 6.20 |
| 0.6 | 1.02 | 0.6 | 10.37 |
| 0.9 | 1.18 | 0.9 | 13.63 |

## 3. Results and discussions

Before presenting the results of the sandwich nanostructure, and for the sake of validation of our computational approach, we calculated the properties of ZnO and TiO$_2$ in bulk systems. Within the above presented computational scheme, the lattice parameters computed for anatase TiO$_2$ in bulk structure are $a = 3.78$ Å and $c = 9.61$ Å in very well agreement with experiment ($a = 3.78$ Å, $c = 9.50$ Å) [35] and other theoretical calculations [36]. To address the accuracy of our method, we calculated structural parameters for TiO$_2$ in rutile phase (with the same calculation parameters described above). The results are reported in Table 2 and in accordance with the previously published ones [36, 37, 27]. We found bulk anatase gives lower binding energy than that of rutile phase: our calculations give a notable similitude as compared to the experimental results. In fact, we found bulk rutile more stable than anatase phase of about 24 meV (0.55 Kcal/mol), which agrees with the experimental value of ~ 1.2 Kcal/mol [38].

**Table 2.** Calculated parameters for ZnO, anatase TiO$_2$, and rutile TiO$_2$ in bulk structures at GGA (GGA+U) level: $a$ and $c$ are relaxed lattice parameters, $E_g$ is the energy band gap

|           | ZnO         | anatase TiO$_2$ | rutile TiO$_2$ |
|-----------|-------------|-----------------|----------------|
| $a$ (Å)   | 3.23 (3.22) | 3.78 (3.74)     | 4.59 (4.52)    |
| $c$ (Å)   | 5.27 (5.17) | 9.61 (9.63)     | 2.97 (2.98)    |
| $E_g$ (eV)| 0.71 (3.42) | 2.03 (3.16)     | 1.82 (2.94)    |

It is well known that DFT-GGA describes the structural relaxation quite accurately, but fails to reproduce the correct alignment in the gap region, which leads to an underestimation of the band gap in the case of strongly correlated systems. To deal with such inconvenience in the interface study, we performed GGA+U [39] calculations for ZnO wurtzite and TiO$_2$ in both anatase and rutile phases. After several trials, we identified the values of the effective potential, $U$, for which the experimental band gap and lattice parameters were successfully reproduced ¶. We found a suitable $U_{O2p}$ = 2.4 eV for the oxygen 2$p$ orbitals, and $U_{Zn3d}$ = 8.0 eV for 3$d$ orbitals of zinc, whereas for the 3$d$ orbitals of titanium, the effective potential shift $U_{Ti3d}$ = 1.4 eV. These parameters reduce the interaction between VB and CB, and are used for the interface band structure calculations. The optimized lattice parameters for ZnO and TiO$_2$ in bulk systems at GGA+U level are reported in Table 2 (the values between brackets). The direct gap found in rutile is about 2.94 eV, while for the indirect gap in anatase phase is of 3.16 eV, in accordance with experiment [40]. The corresponding band dispersions along the high symmetry points, if compared to the ones obtained with GGA, are almost identical. The VB width is slightly increased of 0.12 (0.05) eV for rutile (anatase). The O$_{2s}$ states are shifted up (around 1.5 − 1.8 eV) in both anatase and rutile structure, while in the case of ZnO, the 3$d$ electrons recover the correct alignments and found at ∼ 7 eV below the valence band maximum (VBM), approaching the experimental [26] and theoretical values obtained by employing hybrid functionals [43]. Furthermore, GGA+U calculations give also more accurate value of the relative energy ∼ 55 meV (∼ 1.27 Kcal/mol), and very close to the experimental value [38] in which rutile phase is the most-*likely* stable under thermodynamic equilibrium conditions.

*3.1. Interface relaxations*

Our final relaxed double heterostructure is depicted in Fig. 2. Both types of configurations (left and right interfaces) are characterized by two kinds of relaxation that have different interface hybridization. For example, at the left interface, the oxygen atom which lies at the core dislocation [labeled O1 in Figs. 2(b) and 3] relax inwards in both equivalent sites [surrounded by a circle in Fig. 2(b)], influencing thereby the connected atoms (Zn4 and Zn5 are pushed back, see Fig. 3). In contrast, at the right interface, the Ti atom located in front of Zn11 [labeled Ti11 in Figs. 2(b) and 3] relax outward with respect to titane planes causing bond distortions throughout the lattice (the dimer Zn11−O11 pushed inwards, see Fig. 3). Atoms of $TiO_2$ in both sides of the surface undergo collective relaxation, whereas the ones at the interfaces form with those of ZnO a rhombohedron-*like* lattices (at the left heterojunction) and non-regular hexagonal-*like* lattices (at the right heterojunction) [see Figs. 2(b) and 3]. Surface relaxations of the outer layers are similar in both sides, identical to those characterizing the relaxation of (101) anatase $TiO_2$ [27, 28], and the atoms therein (bulk-*like*) are not affected by distortions that occur at interfaces.

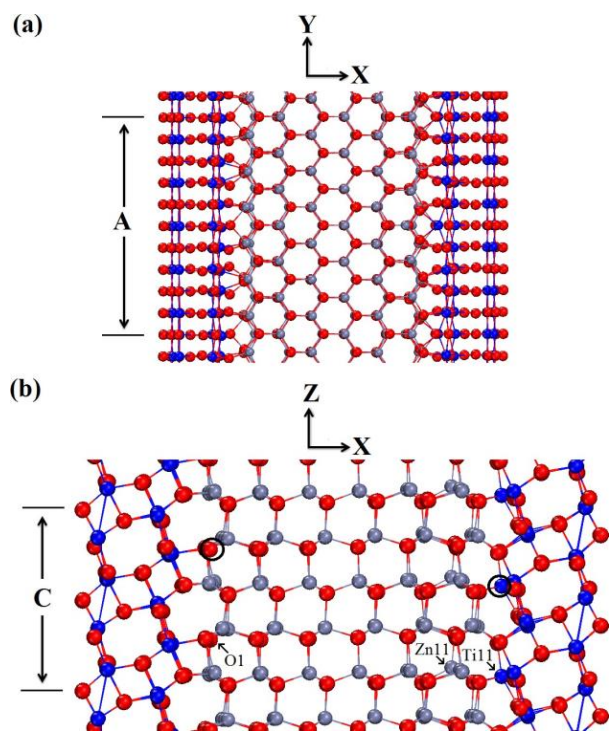

**Figure 2.** Relaxed structure of the sandwich system $TiO_2$/ZnO/$TiO_2$: projection in (*XY*) plane (a), projection in (*XZ*) plane (b). Oxygen, zinc, and titanium are represented by red, gray, and blue balls, respectively.

If comparing atomic bond lengths in the relaxed system to those in the respective ideal-bulk structure of $TiO_2$ and ZnO, one may note that these lengths are less distorted in the right interface than those of the left one (see Table 3). Hence, the resulting total energy with GGA (GGA+U) of the right interface lower than that of the left one with about 2.6 (1.6) eV [+]. Some interface bond lengths and interatomic distances as labeled in Fig. 3 are summarized in Table 3. The lengths of the dimers $O_1-Zn_{1,4,5}$ at the left interface are contracted with respect to ZnO bulk value (~ 1.98 Å [32]), whereas, $O_2-Zn_{1,2}$ and $O_3-Zn_{1,3}$ are larger. In contrast, at the right one, small bond distortions have been noticed

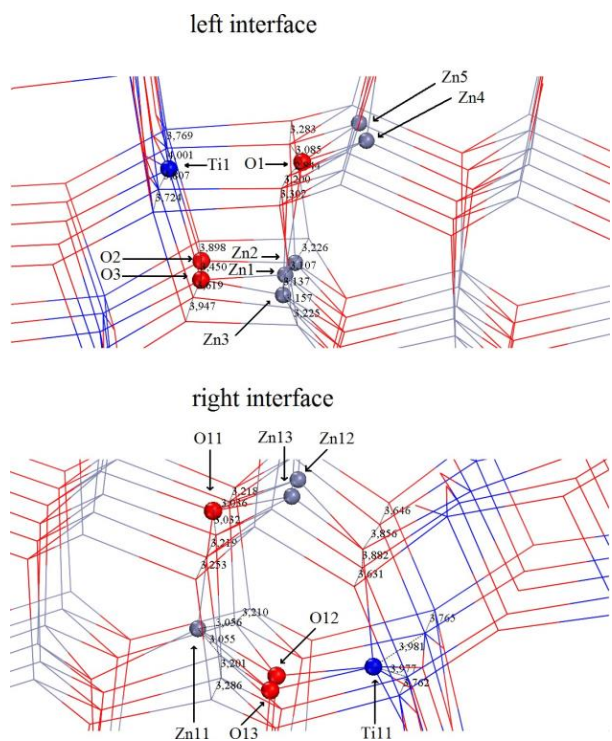

**Figure 3.** Interface hybridizations and atomic bonding at the two interfaces of the relaxed $TiO_2/ZnO/TiO_2$ system.

(compare $O_{11}-Zn_{11,12,13}$ and $Zn_{11}-O_{12,13}$ lengths with 1.98 Å). It is worth noting that the left heterojuction presents two dangling bonds (O1 and its equivalent site), whereas in the right interface no dangling bonds were observed (see Fig. 3). Such structural changes induced by misfit dislocations entail the difference in total energy between the two interfaces, giving rise to an offset in the alignment of the energy levels between the two junctions as will presently appear below.

**Table 3.** Interatomic distances at the interfaces of the relaxed $TiO_2/ZnO/TiO_2$ system. Atom labels refer to Fig. 3

| atomic label | length (Å) | atomic label | length (Å) |
| --- | --- | --- | --- |
| O1–Zn1 | 1.93 | O11–Zn11 | 1.94 |
| O1–Zn4 | 1.90 | O11–Zn12 | 2.04 |
| O1–Zn5 | 1.91 | O11–Zn13 | 2.04 |
| Zn1–O2 | 2.33 | Zn11–O12 | 1.98 |
| Zn1–O3 | 2.11 | Zn11–O13 | 1.98 |
| Zn2–O2 | 2.00 | Ti11–O12 | 2.08 |
| Zn3–O3 | 2.08 | Ti11–O13 | 2.08 |
| Ti1–O1 | 2.77 | O12–O13 | 2.54 |

*3.2. Electronic properties*

One of the queries, not yet properly explored from a simulation point of view that we need to clarify is to overcome the band gap error in the interface band structure calculations. For this reason, we performed GGA+U simulations using the potential shift parameters carefully checked and described above. The band structure with the corresponding density of states (DOS) are reported in Fig. 4. The improvements made with respect to GGA results include: i) the direct energy gap is found more than twice larger [around 1.03 (2.53) eV with GGA (GGA+U)], ii) the electronic wave functions at the VB edges are localized, iii) the conduction band is mainly maintained by $Ti_{3d}$ orbitals in which the 4*s* orbitals of zinc are pushed up [see Fig. 4(b)]. The direct gap calculated from the band diagram of Fig. 4(a) is very close to the optical gap found experimentally by Kayaci *et al.* [44] ($\sim 2.45$ eV).

One of the different results of the sandwich system, if compared to a single heterostructure, can be observed from the DOS reported in Fig. 4(a). The trapped or interface states at the vicinity of VB are virtually eliminated, which states whose effects are detrimental for the efficiency of the DSSC. Such behavior has been experimentally verified by many authors [11, 14, 16, 17, 18, 19, 20, 45, 46, 47, 48] but with inhomogeneous efficiencies. One can also notice that the ZnO CB density shifted to higher energies by the larger density of 3d orbitals of titanium [see Fig. 4(b)], resulting in an increase of the excited electron lifetime. We have to underline that for an inverted sandwich structure, i.e., $ZnO/TiO_2/ZnO$, the electronic and optical properties may behave differently due to the lower density of states in CB of $TiO_2$ where the thickness of the two different materials affects significantly the band alignments as it was reported in Refs. [44, 47, 48]. Furthermore, it can be explained by the different electrostatic interaction at the semiconductor surface [44, 49] and, unlike ZnO, in $TiO_2$ the charge recombination is very slow. Němec *et al.* [49] attribute the different charge transport and recombination in the two semiconductors to the screening of the electrostatic interaction in $TiO_2$ due to its high dielectric permittivity.

Understanding the effect of atomic bonding and strains on the electronic states at the edge of the VB and CB, the spatial distribution (wave functions) of some occupied and unoccupied states as obtained within GGA+U functional are plotted at Gamma- point [**G** as labeled in Fig. 4(a)] and illustrated in Fig. 5. One can see in clear the role of bonding and local atomic distortions on the overlap of the quantum states. In particular, the highest occupied state (HO) and (HO-1) are defect states that arise mainly from $O1_{2p_{z,x}}$ and its likely-equivalent atom, respectively (localized on the core dislocations at the edge of the left interface). Whereas the lowest unoccupied state (LU) and (LU+1) derived mainly from $Ti_{3d}$ orbitals of titane atoms of the left heterojunction. These interface states located at $\sim 2.5$ eV above VBM are associated with the broad visible emission centered at $\sim 2.4-2.5$ eV found in the experimental work of Kayaci *et al.* [44] (which they attributed to bulk grain transition and oxygen vacancy). The above findings evidence that the conduction band potential of the left interface is more negative than that of the right one. The conduction band of $TiO_2$ in both sides of the double interface is more negative than that of ZnO in accordance with experimental results reported in Refs. [18, 50, 51, 44] and at variance to what have been suggested in other works [21, 45].

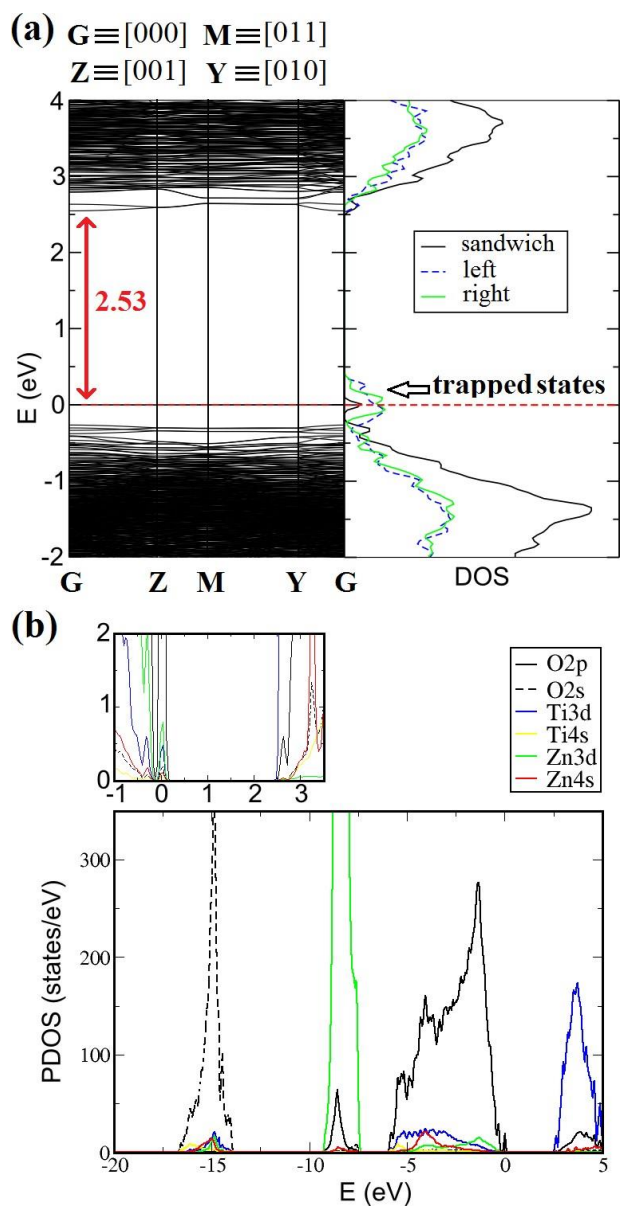

**Figure 4.** Band structure plot along high symmetry points together with the total density of states, DOS, at GGA+U level, (the comparison between different densities is made by aligning their *s* and *d* states) (a), the projected density of states, PDOS, (b). The zero of energy is set to the VBM of the sandwich structure.

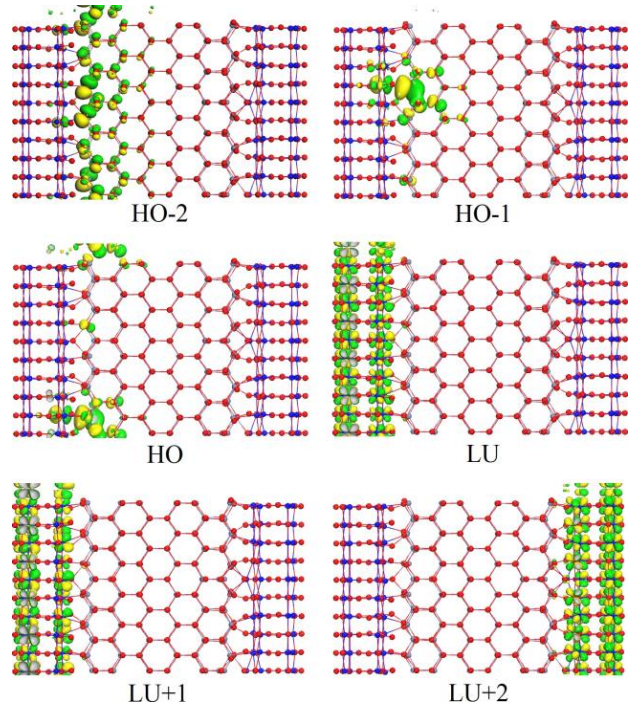

**Figure 5.** Isosurface plots of the spatial distribution of the wave functions around VB and CB at Gamma-point as obtained for $TiO_2/ZnO/TiO_2$ double heterostructure at GGA+U level. Isovalue of 0.05 e/Å$^3$.

### 3.3. Interface band offsets

To dispel doubts over the above issue and, for a better and clear description of the relative positions of energy levels at the interfaces [#], a lineup of the average of the electrostatic potential between the two materials is required. To define band offsets: valence band offset (VBO) and conduction band offset (CBO), we employ the method described in Refs. [52, 53, 54]. These are calculated according to the following relation: VBO (CBO)= $\Delta E_v(\Delta E_c) + \Delta V$, where $\Delta E_v$ ($\Delta E_c$) is the so-called *band-structure* term, which refers to the difference between the top (bottom) of the valence (conduction) bands as obtained from two independent bulk band structure calculations. $\Delta V$ is the lineup of the average of the electrostatic potential through heterojunctions. The quantity $\Delta V$ contains all interface effects that result from electronic charge transfer after interfacial hybridization.

The plot of the average of the electrostatic potential is illustrated in Fig. 6(a). In this case, only the VBO is calculated from the double-macroscopic average technique, and by using the above formula, we deduce CBO by adding the experimental gap of the materials constituting our sandwich structure.

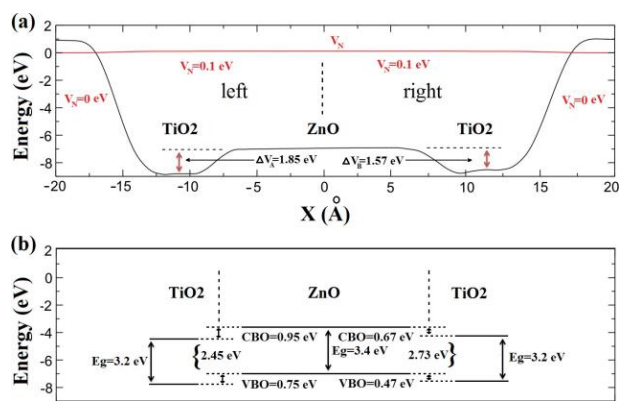

**Figure 6.** Diagram of the average of the Hartree potential and the neutral potential $V_N$ as obtained using the double-macroscopic average technique (a). Schematic representation of the valence band offsets (VBO) and conduction band offsets (CBO) for $TiO_2/ZnO/TiO_2$ interfaces (b). The energy gap, $E_g$, refers to the experimental gap of bulk systems.

The computed values are depicted schematically in Fig. 6(b). They evidence a heterojunction type II, in agreement with experimental observations [18, 44, 50, 51, 55, 56]. The larger energy gradient between CB (VB) of ZnO and CB (VB) of $TiO_2$ favors the transfer of the excited electrons (active holes) from ZnO ($TiO_2$) to $TiO_2$ (ZnO). The striking feature regards the difference in band offsets between the left and right interfaces, in which the conduction band potential of the left side is more negative with about ∼ 0.23 −0.28 eV, and thereby leading to an electron accumulation in the CB of the left shell of $TiO_2$ rather than the right one. The VBO (CBO) is found to be 0.75 (0.95) eV and 0.47 (0.67) eV for the left and the right interface, respectively (see Fig. 6), predicting a more enhanced charge transfer/separation for the case of the left interface structure. The above result points to challenges in fabricating a double heterojunction with desirable interfacial structures, since inappropriate structures of the interface can lower the desired properties of the hybrid materials [18]. This result also reveals to what extent the potential of VB and CB can be modified by interfacial effects, and may explain why there have been controversial published results [18, 21, 45, 48, 50, 51] as well as on the energy band alignments and on the effect of ZnO coating layer on the device performance of a DSSC.

It is worthwhile to note that the minimal gap calculated from the double- macroscopic average technique [∼ 2.45 eV from Fig. 6(b)] close to the gap value given by GGA+U calculations [∼ 2.53 eV, see Fig. 4(a)], and in very good agreement with the experimental results of Ref. [44] (∼ 2.45). Moreover, the LU+2 state maintained by $Ti_{3d}$ of the right side of $TiO_2/ZnO/TiO_2$ system (see Fig. 5) is found at 2.71 eV above VBM, [HO→(LU+2)], nearly to what we estimated from the gap region of the right interface ∼ 2.73 eV [see Fig. 6(b)]. This transition can be associated to that observed in the experimental work of Kayaci *et al.* [44] (located at ∼ 2.8 eV). The above remarks demonstrate the accuracy of our GGA+U results and validate the double-macroscopic average technique for the determination of the band offsets in semiconductor/semiconductor interfaces.

In the experimental results published so far, several works [11, 14, 18, 50, 56] pointed out qualitatively the band alignments between the two materials but does not allow a direct comparison. To the best of our knowledge, some values on ZnO/anatase-type $TiO_2$ heterojunction band offsets have been obtained by Ran Zhao *et al.* [51], that are of 0.2 (0.6) eV for CBO (VBO), quantitatively closer to what we found for the right interface. Similar results were found in Ref. [55] in which the CBO estimated to be 0.44 eV for 0.7 nm thick ZnO that decreases upon thickness expansion of ZnO. While measurements on wurtzite ZnO/rutile-type $TiO_2$ by x-ray photoelectron spectroscopy (XPS) [50] evaluate VBO (CBO) to be $0.14 \pm 0.05$ ($0.45 \pm 0.05$) eV.

## 4. Conclusions

In summary, an extensive study on the nature of the interface between ZnO (wurtzite) and $TiO_2$ (anatase) has been presented by means of DFT(+U) simulations. Our structural analysis allowed us to obtain; an accordance between the atomic planes of the two materials in the $TiO_{2-[101]} \| ZnO_{[0001]}$ interface direction, and a misfit dislocation in the perpendicular direction $TiO_{2-[010]} \| ZnO_{[11\bar{2}0]}$, which may explain the origin of diffusion and oxygen vacancy reported in some experimental observations [11, 14, 19, 20, 44]. Using the ground state energy calculations, we found two most stable configurations that have different interface hybridization, from which our double heterostructure $TiO_2/ZnO/TiO_2$ has been constructed. Besides, the non-equivalence of the atomic environment and the presence of more dangling bonds in the left interface makes it less stable than the right one. The study of the electronic properties at GGA+U level shows that valence band is maintained by $O_{2p}$ orbitals of ZnO, whereas conduction band arises mainly from $3d$ orbitals of titanium in which $Zn_{4s}$ were slightly pushed up. Furthermore, by using the double-macroscopic average technique, we evidence a heterojunction type II but we observed a worthy difference in energy levels between the two interfaces (left and right), resulting in an electron accumulation in the CB of the left shell of $TiO_2$ rather than the right one. Such a double interface system demonstrates an efficient charge separation and increasing of excited electron lifetime, in particular, when $TiO_2$ is used as a shell with a suitable thickness [44]. In practical point of view, our study allows us to understand the mechanism of the oxide hetero-interfaces and the origin of some optical transitions which can open new opportunities for a better use in photocatalytic and photovoltaic based devices.

## Acknowledgments

The authors are grateful to the South African Center for High Performance Computing (CHPC), Projet N. MATS0868 and CINECA Award N. HP10CLG9UX (Italy) for the availability of high performance computing resources and support.

## Footnotes

‡ These are lattice parameters optimized for bulk calculations.
§ In fact, these atomic layers are spaced by the lattice fringes characterizing (112) planes of anatase $(10\bar{1})$ or (101) surfaces of $TiO_2$.

\* Our theoretical deductions are in agreement with the HRTEM image of the work done by Matt Law *et al.* [11].

¶ It should be known that the appropriate choice of *U* depends on several parameters namely: the DFT code used, the choice of basis sets, pseudopotentials, and K-points mesh [42].

⁺ It should be noted here that we also made relaxations for the two interfaces separately, i.e., for $TiO_2$/ZnO (left) and ZnO/$TiO_2$ (right).

# Even if we can have an idea about band alignments around valence and conduction bands from results obtained with GGA+U functional, it is necessary to confirm or refute it by using another method namely double-macroscopic average technique [53, 54, 25].